\documentclass{ws-mpla}

\usepackage{parskip}
\usepackage{graphicx}
\usepackage{hyperref}

\usepackage{bm}
\usepackage{amsbsy}

\usepackage{aas_macros}
\usepackage[super,compress]{cite}
\usepackage[percent]{overpic}

\newcommand{\pb}{\textsc{Polarbear}}

\title{THE COSMIC MICROWAVE BACKGROUND AND PSEUDO-NAMBU-GOLDSTONE BOSONS: SEARCHING FOR LORENTZ VIOLATIONS IN THE COSMOS}
\author{DAVID LEON}
\address{Department of Physics, University of California, San Diego, 9500 Gilman Drive, La Jolla, CA 92093-0424, USA\\dleon@physics.ucsd.edu}
\author{JONATHAN KAUFMAN, BRIAN KEATING}
\address{Department of Physics, University of California, San Diego, 9500 Gilman Drive, La Jolla, CA 92093-0424, USA}
\author{MATTHEW MEWES}
\address{California Polytechnic State University, San Luis Obispo, CA 93407, USA}
\date{}

\begin{document}

\markboth{Leon, Keating, Kaufman, Mewes}{The CMB and PNGBs: Searching for Lorentz Violations}

\catchline{}{}{}{}{}

\maketitle

\begin{abstract}

One of the most powerful probes of new physics is the polarized Cosmic Microwave Background (CMB). The detection of a nonzero polarization angle rotation between the CMB surface of last scattering and today could provide evidence of Lorentz-violating physics. The purpose of this paper is twofold. First we review one popular mechanism for polarization rotation of CMB photons: the pseudo-Nambu-Goldstone boson. Second, we propose a method to use the \pb\ experiment to constrain Lorentz-violating physics in the context of the Standard-Model Extension, a framework to standardize a large class of potential Lorentz-violating terms in particle physics.

\end{abstract}


\ccode{PACS numbers:}

\section{Introduction}

The principle of Lorentz invariance undergirds our two most fundamental theories of physics: the standard model of particle physics and general relativity. While these theories have been extraordinarily successful at describing nature, we expect new physics to emerge at high energies. Since we have limited ability to directly probe the regimes where both quantum mechanics and relativity dominate, we can instead search for small deviations in our low-energy theories such as a departure from Lorentz symmetry.

One promising way to search for Lorentz violations is to study linearly polarized light as it propagates through space. From rotational symmetry, it is expected that the polarization plane of a linearly polarized photon should not change as it propagates through empty space, so any rotation of the polarization would indicate some physical interaction that violates Lorentz invariance.

Although there are a number of potential theories that could cause such a rotation, one popular mechanism for achieving this effect is the introduction of an additional scalar field to the standard model. The pseudo-Nambu-Goldstone boson (PNGB) has a number of attractive qualities, including the ability to conserve an otherwise broken global symmetry and to naturally explain the connection between symmetry-breaking effects over many orders of magnitude in energy. It is also a potential candidate to account for the observed dark energy or dark matter densities.  Since PNGBs generically introduce a Lorentz-violating polarization angle rotation via Chern-Simons coupling to electromagnetism, we also have a promising avenue to search for their existence.\cite{sikivie2008axion}$^,$\cite{ratra1988scalar}$^,$\cite{rosenberg2014axion}$^,$\cite{ng2001pngb}$^,$\cite{ferreira2014axion}

Of course this is only one possible mechanism so we should also consider ways to classify more general Lorentz violations, which is the aim of the Standard-Model Extension (SME) framework. In particular the CMB is an excellent probe of Lorentz violations because small effects can accumulate over cosmological distances. Using observations from the \pb\ experiment\cite{PB1Bmodes} we propose a method to constrain a subset of the SME parameters using the three CMB patches observed by \pb.

The paper is organized as follows. Sections \ref{sec:scales} through \ref{sec:birefringence} are a brief review of a general PNGB and its evolution through the universe, including specific examples like the QCD axion and quintessence; a candidate for dark energy. Sections \ref{sec:sme} through \ref{sec:polarbear} deal with the SME and show how cosmic polarization rotation angle measurements can be used to constrain SME parameters.

\section{Energy Scales in Field Theory}
\label{sec:scales}

There is a general heuristic in field theory that can be expressed in imprecise terms as: ``mass terms in the action correspond to energy scales of the relevant physics.''

This statement is tantamount to dimensional analysis. When using natural units that set $c=1$ and $\hbar=1$ any quantity can be written in terms of powers of a single unit, usually in terms of eV, sometimes in terms of mass. In the path integral approach to quantum mechanics, the action appears in an exponential $e^{iS}$ so dimensional analysis tells us that the action must be dimensionless. We can then examine the definiton of the action
\begin{equation} S=\int d^4x \mathcal{L}(\phi) . \end{equation}
From this one can conclude that, in 4D space, the Lagrangian has units of [distance]$^{-4}$, or equivalently [mass]$^4$, in order to make the action unitless. Quantum fields also have mass dimensions. For example, in order for the kinetic term of the scalar field, $\frac{1}{2}\partial_\mu \phi \partial^\mu \phi$, to have dimension 4, and noting that a derivative has mass dimension 1, then it is necessary for the field $\phi$ to have dimension 1. A consequence of this is that a term like $\phi^2$ has dimension 2 and therefore needs a dimensionful coupling in order to be in the Lagrangian, for example: $m^2\phi^2$, where $m$ is a constant with mass dimension 1.

A more complicated example is the Chern-Simons term between a scalar field and a field strength tensor: $\phi F^{\mu\nu} \widetilde{F}_{\mu\nu}$. In electromagnetism the field strength tensor $F_{\mu\nu}$ is constructed from derivatives and the vector potential (mass dimension 1)
\begin{equation} F_{\mu\nu} = \partial_\mu A_\nu - \partial_\nu A_\mu . \end{equation}
Therefore $F_{\mu\nu}$ has dimension 2, and the overall Chern-Simons term has dimension 5, so it must be accompanied by a coupling of dimension -1. Therefore something like a constant $1/M$ for a Lagrangian term of the form
\begin{equation} \frac{1}{M} \phi F^{\mu\nu} \widetilde{F}_{\mu\nu} \end{equation}
would be a candidate ansatz. Considering a theory that contains terms with dimensionful constants, a reasonable question to ask is whether the value of that constant is an arbitrary parameter of the theory or whether it traces new physics at roughly that scale, which is not yet understood. Take for example the masses of the $W$ and $Z$ bosons. They are not arbitrary parameters but they come from electroweak-symmetry breaking. Furthermore both of these particles and the Higgs boson all have masses of around 100 GeV, which is on the order of the electroweak scale $\sim 200$ GeV.

Similarly the pion of mass $\sim 130$ MeV and the proton/neutron masses $\sim 1000$ MeV are all around the same order of magnitude as the QCD scale $\sim 200$ MeV.

This reasoning also leads us to believe there is some, as yet unknown, physics at the Planck scale since general relativity has a dimensionful parameter of its own: $G = \frac{1}{M_P^2}$.

This energy-scale approach is especially relevant to cosmology because it deals with the evolution of the universe, which spans some 30 orders of magnitude in temperature, with the temperature going down as the universe ages. Even an order of magnitude approach to cosmological fields should lend us considerable insight. So with this heuristic in mind, we now take a look at the PNGB.

\section{Mass of the Pseudo-Nambu-Goldstone Boson}

If a scalar field theory with multiple degrees of freedom is invariant under some symmetry, that symmetry can be spontaneously broken via the Higgs mechanism. To understand how this happens, first consider an example with a single scalar field.

\subsection{Spontaneous Symmetry Breaking}

A single (real) scalar field, which has only one degree of freedom, has a Lagrangian of the form
\begin{equation} \mathcal{L} = \frac{1}{2}\partial_\mu \phi \partial^\mu \phi - V(\phi) . \end{equation}
The lowest-energy state is simply a constant value $\phi = \phi_0$. The derivative terms are zero for a constant, so its value is determined by the minimum of the potential, $V(\phi_0)$. Including a mass term or even a higher-order quartic term results in a potential
\begin{equation} V(\phi) = \frac{1}{2}\mu^2 \phi^2 + \frac{1}{4}\lambda \phi^4 . \end{equation}
This potential, as well as the full Lagrangian, has a discrete sign symmetry. Replacing $\phi$ with $-\phi$, the Lagrangian is unchanged because only even powers of $\phi$ appear. As long as the parameters $\mu^2$ and $\lambda$ are positive, then the minimum of this potential is $\phi_0 = 0$. However, in field theory it is usually the case that ``constants'' in the Lagrangian are not truly constant. Due to the process of renormalization, the coupling ``constants'' can actually depend on energy scale. For example, asymptotic freedom in QCD is related to the statement that the strong fine-structure parameter $\alpha_s$ is large at low energies, but becomes asymptotically smaller at larger energies.

\begin{figure}
  \centering
  \includegraphics[width=0.49\textwidth]{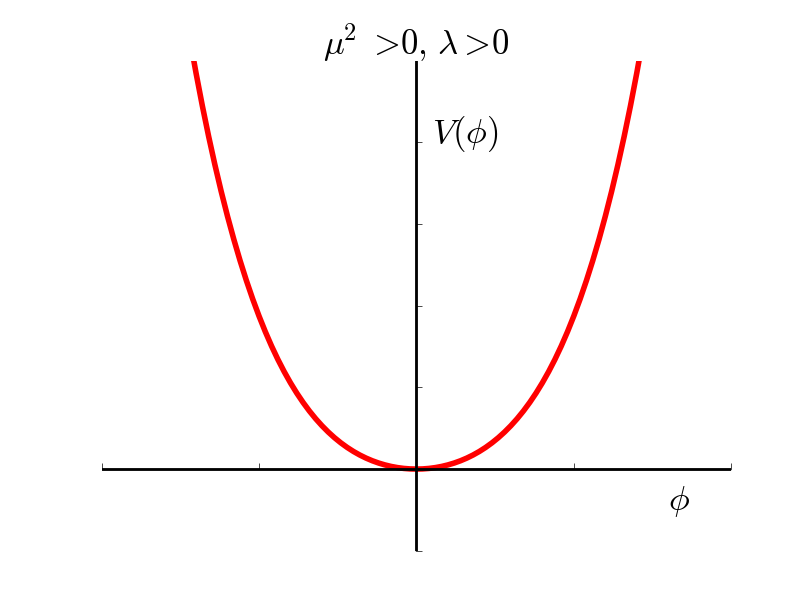}
  \includegraphics[width=0.49\textwidth]{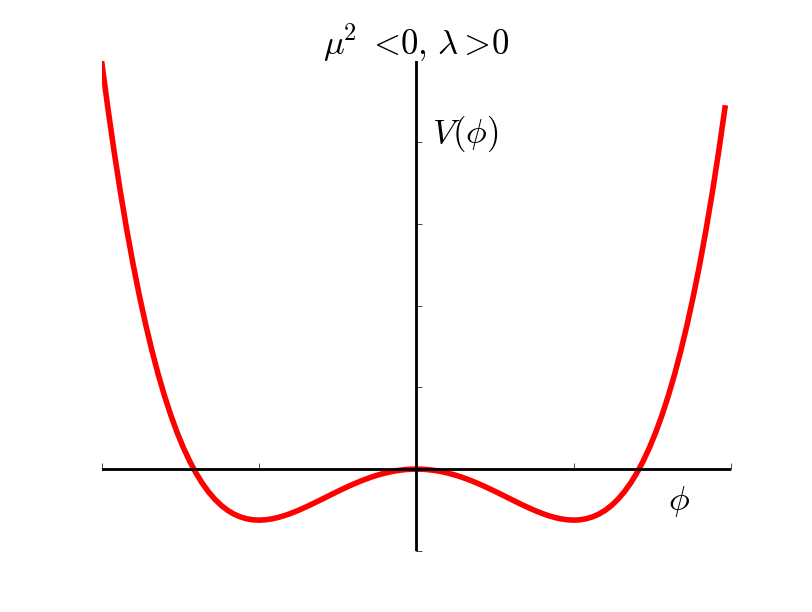}
  \caption{Scalar potential for positive and negative mass parameters.\label{fig:higgs}}
\end{figure}

If, at some point in the evolution of the universe, the temperature-dependent parameter $\mu^2 = \mu^2(T)$ changes from a positive value to a negative value, then the minimum of the potential will no longer be $\phi_0 = 0$. Figure~\ref{fig:higgs} shows plots of the potential $V(\phi)$ for two cases $\mu^2>0$ and $\mu^2<0$. In the second case there are two minima at
\begin{equation} \phi = \pm v = \pm \sqrt{\frac{-\mu^2}{\lambda}} . \end{equation}

After transitioning to $\mu^2<0$ the Lagrangian is still symmetric but expansion of the Lagrangian about the ground state, in terms of $\phi - \phi_0$, is no longer symmetric. This phenomemon is called ``spontaneous symmetry breaking'' because, while the symmetry of the underlying Lagrangian is not broken, symmetry of the ground state \textit{is} broken. In this example the symmetry was discrete but in a model with more degrees of freedom, spontaneous symmetry breaking of a continuous symmetry will occur.

\subsection{Complex Scalar Field}

Next consider a complex scalar field $\psi = \psi_1 + i\psi_2$ with Lagrangian
\begin{equation} \mathcal{L} = \partial_\mu \psi^\dagger \partial^\mu \psi - V(\psi) . \end{equation}
This field will have a similar potential
\begin{equation} V(\psi) = \mu^2\psi^\dagger \psi + \lambda (\psi^\dagger \psi)^2 . \end{equation}
Under a spontaneous-symmetry-breaking event as described above, $\mu^2$ will become negative and the potential will take on the new form
\begin{equation} V(\psi) = \lambda(\psi^\dagger\psi - v^2)^2 . \end{equation}
In terms of the component fields $\psi_1$ and $\psi_2$ this is
\begin{equation} V(\psi_1,\psi_2) = \lambda(\psi_1^2 + \psi_2^2 - v^2)^2 . \end{equation}
This can be parameterized in terms of two different real scalar fields $H$ and $\phi$
\begin{equation} \psi = He^{i\phi/f} , \end{equation}
so that
\begin{equation} \label{eq:psi1} \psi_1 = H\cos(\phi/f) , \end{equation} 
\begin{equation} \psi_2 = H\sin(\phi/f) . \end{equation}

\begin{figure}
  \centering
  \begin{overpic}[width=\textwidth]{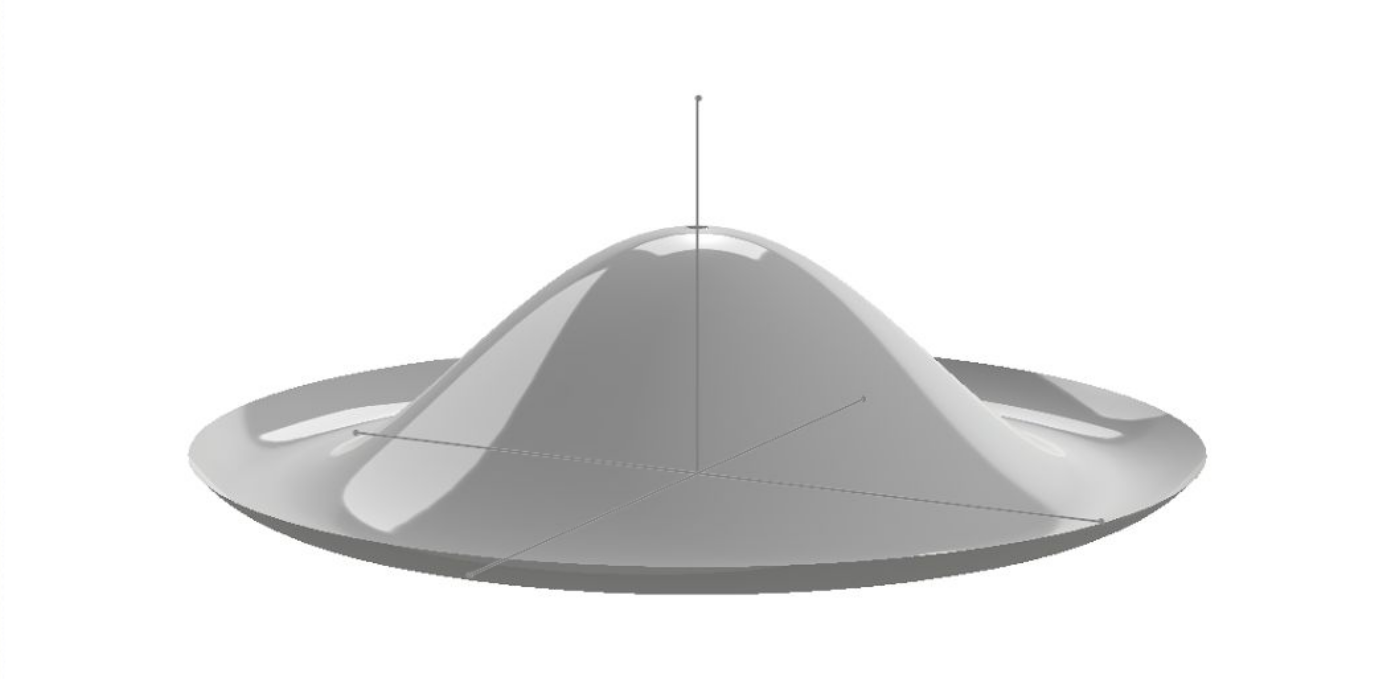}
    \put (28,4) {$\psi_1$}
    \put (85,9) {$\psi_2$}
    \put (47,45) {$V(\psi)$}
  \end{overpic}
  \caption{Potential of a Nambu-Goldstone boson.\label{fig:GoldstonePotential} }
\end{figure}

This yields a potential only dependent on the longitudinal degree of freedom $H$, as seen in Figure~\ref{fig:GoldstonePotential}
\begin{equation} V(H,\phi) = V(H) = \lambda(H^2 - v^2)^2 . \end{equation}
This result, where the potential depends on some degrees of freedom but not others, is a general feature of spontaneous-symmetry-breaking events. The degrees of freedom that do not appear in the potential after symmetry breaking are called ``Nambu-Goldstone bosons,'' and the lack of a potential means they are massless by definition. In this example $\phi$ is the Nambu-Goldstone boson. Additionally they possess shift symmetry in which a change in the value of $\phi$ by $2\pi f$ corresponds to the same physical system
\begin{equation} He^{i\phi/f} \to He^{i(\phi+2\pi f)/f} = He^{i\phi/f} . \end{equation}
This new parameter $f$ is included because if $\phi$ is interpreted as a scalar field, dimensional analysis tells us that in order to have it inside the argument of an exponential it must be divided by a coupling with dimension 1. In the spirit of the previous section, it is reasonable to expect that the value of the coupling $f$ will be on the same order of magnitude as the other scales in this scenario like the vacuum expectation value $f\sim v$.

This is the situation with only a scalar field and a spontaneous-symmetry-breaking event. But PNGBs have a second relevant temperature scale when the shift symmetry of $\phi$ is broken. Next consider a case where the scalar is coupled to either some other field, or collection of fields, or constant term generated by other fields, so that the potential is
\begin{equation} V(\psi,\Psi) = \lambda(\psi^\dagger\psi - v^2) - \beta\frac{\psi_1}{f} \Psi . \end{equation}
Where $\beta$ is a dimensionless coupling of order unity. In this case, not only will the field undergo spontaneous symmetry breaking, but it will experience explicit symmetry breaking because the $\psi_1 \Psi$ term does not respect the U(1) symmetry of $\psi$: the Lagrangian is not invariant under the global transformation $\psi \rightarrow \psi' = e^{i\theta} \psi$.

\begin{figure}
  \centering
  \begin{overpic}[width=\textwidth]{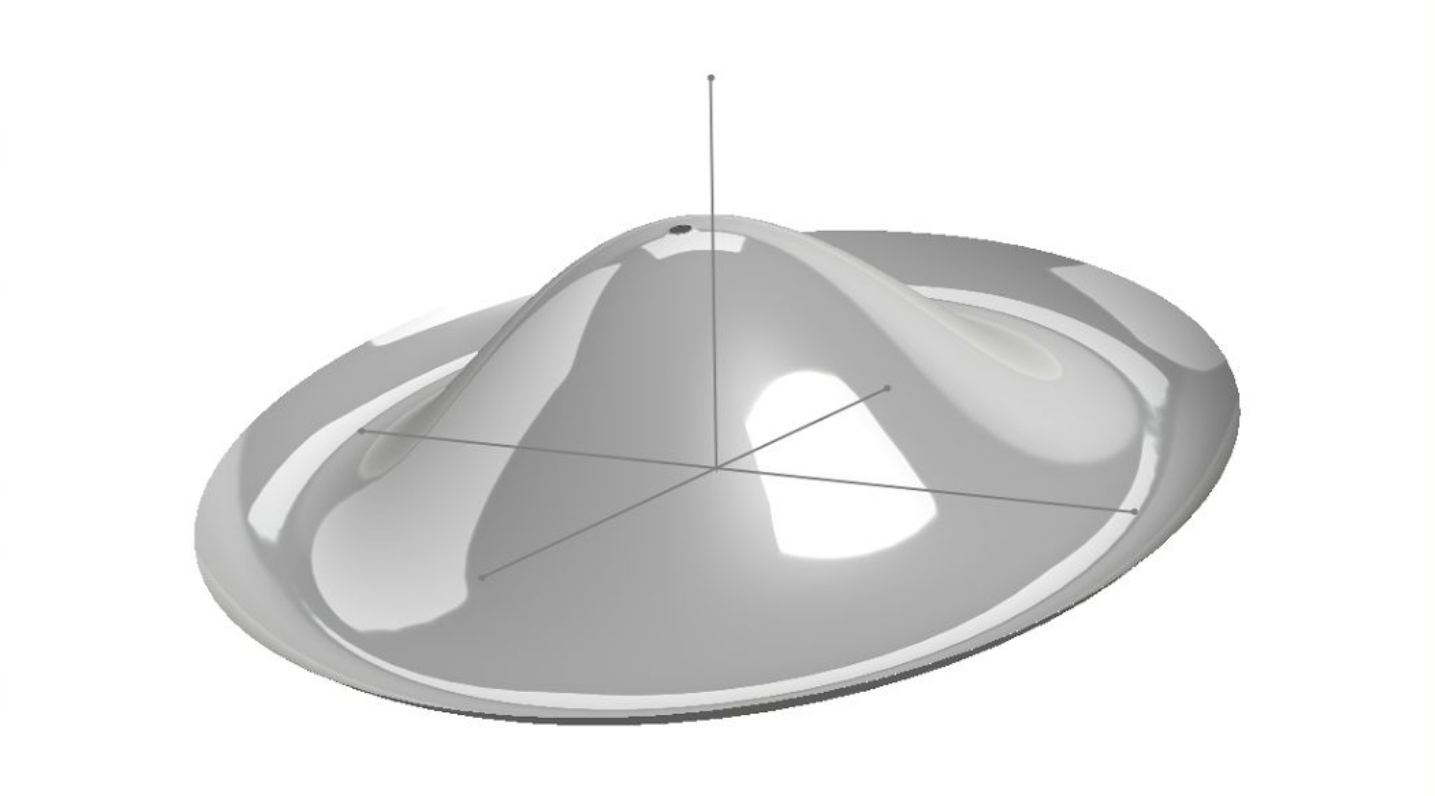}
    \put (16,7) {$\psi_1$}
    \put (85,18) {$\psi_2$}
    \put (47,53) {$V(\psi)$}
  \end{overpic}
  \caption{Potential of a pseudo-Nambu-Goldstone boson.\label{fig:PNGBpotential} }
\end{figure}

In this example $\Psi$ represents whatever physics is explicitly breaking the symmetry. This is model dependent, but by dimensional analysis it must have mass dimension 4 regardless of the model. The potential can be seen in Figure~\ref{fig:PNGBpotential}

Now when reparameterizing the $\psi$ field, the potential becomes
\begin{equation} V(\psi,\Psi) = \lambda(H^2 - v^2) - \frac{\beta}{f}H\cos(\phi/f) \Psi . \end{equation}
In the low-energy limit, $\phi \ll f$, the quadratic term from the cosine in Eq.~(\ref{eq:psi1}) will dominate and the potential for the PNGB is
\begin{equation} V(\phi) \approx \frac{\beta}{2f^3}H\phi^2\Psi . \end{equation}
Additionally, in the low-energy limit, the $H$ field will be near its equilibrium value $v$, which should be around the same size as $f$. The potential is therefore
\begin{equation} V(\phi) \approx \frac{1}{2f^2}\phi^2\Psi . \end{equation}
$\Psi$ depends on the model, but the fact that $\Phi$ is a dimensionful term means it should correspond to the temperature scale of the explicit-symmetry-breaking mechanism. Therefore the potential should generally be on the order of
\begin{equation} \label{eq:potentialLambda} V(\phi) \approx \frac{\Lambda^4}{2f^2}\phi^2 , \end{equation} 
where $\Lambda$ now represents the temperature scale of the explicit symmetry breaking. From this scalar mass term, the mass of the PNGB is
\begin{equation} m \approx \frac{\Lambda^2}{f} . \end{equation}
Thus the mass is dependent on the two relevant temperature scales of the PNGB: the spontaneous-symmetry-breaking scale, parameterized by $f$, and the explicit-symmetry-breaking scale, parameterized by $\Lambda$.

\section{Energy Density of the PNGB}

\subsection{Pre-Oscillations}

Now consider the PNGB field in the context of general relativity. The Lagrangian is
\begin{equation} \mathcal{L} = \frac{1}{2}\sqrt{-g} (g^{\mu\nu}\partial_\mu \phi \partial_\nu \phi - m^2 \phi^2 ) . \end{equation}
Focusing only on a spatially uniform field over cosmological scales in the Friedmann-Lema\^{i}tre-Robertson-Walker metric, the Euler-Lagrange equation of motion is
\begin{equation} \frac{d^2\phi}{dt^2} + 3H \frac{d\phi}{dt} + m^2 \phi = 0 . \end{equation}
At this point, the PNGB has first gone through spontaneous symmetry breaking and then explicit symmetry breaking. The equation of motion is of the form of a damped simple harmonic oscillator with a damping term $3H$ and angular frequency $m$. Since $H$ is time dependent the decay envelope will actually be a power law rather than an exponential, but qualitatively the situation is similar. However when the mass first ``turns on'' at the explicit-symmetry-breaking scale $\Lambda$, defined in Eq.~(\ref{eq:potentialLambda}), it is much smaller than $H$. The mass and Hubble parameters are roughly $H \sim \Lambda$, and $m \sim \frac{\Lambda^2}{f}$, therefore their ratio is small $\frac{m}{H} \sim \frac{\Lambda}{f} \ll 1$. This means the mass can be neglected until $H$ drops to a small enough value. Until then the evolution is governed by
\begin{equation} \frac{d^2\phi}{dt^2} + 3H \frac{d\phi}{dt} = 0 . \end{equation}
The solution is simply a constant value $\phi = \left<\phi\right>$. The PNGB sits at the same vacuum expectation value it was given all the way back at spontaneous symmetry breaking. Only once $H$ reaches a small enough value, so that $H\sim m$, do oscillations begin. At this point, the potential energy of the PNGB is about
\begin{equation} \label{eq:potentialPhi} V(\phi) = \frac{1}{2}m^2 \phi^2 \sim m^2 \alpha^2 f^2 . \end{equation} 
Here the vacuum expectation value is $\left<\phi\right> = \alpha f$, where $\alpha$ is a constant of order 1. This is assumed since the initial value of the field should be the same magnitude as the spontaneous-symmetry-breaking scale $\left<\phi\right> \sim f$. At symmetry breaking, $\phi$ takes a value at the bottom of the Higgs potential in the range $-\pi f < \phi < \pi f$. It has no preference for any particular value, so a random value of $\alpha$ in the range from $-\pi$ to $\pi$ will give an expectation value on the order of $f$.

\subsection{Post-Oscillations}

Starting from the initial value of $\phi \sim m^2 \alpha^2 f^2$ from Eq.~(\ref{eq:potentialPhi}), the field then evolves according to the equation of motion stated above, now including mass. The general form of the solution is a sinusoidal oscillation with a decaying power law in time. For example, the solution to the field evolution in a radiation dominated universe is a Bessel function
\begin{equation} \phi(t) \propto t^{-1/4}J_{1/4}(mt) \approx C t^{-3/4}\sin(mt) = C a(t)^{-3/2} \sin(mt) . \end{equation}
Regardless of the time dependence of $H$, the result is that the amplitude of the field oscillations is proportional to the scale factor in this regime
\begin{equation} A_\phi \propto a^{-3/2} . \end{equation}
The energy density of the oscillating field as a function of time is
\begin{equation} \rho = \frac{1}{2}m^2 A_\phi(t)^2 = \frac{1}{2}m^2 A_\phi(t_1)^2\frac{a(t_1)^3}{a(t)^3} , \end{equation}
where $t_1$ is the time at which the oscillations began, at the temperature scale $\Lambda_1 = m$. Finally this allows us to find the energy density today. The initial amplitude of the PNGB oscillations is
\begin{equation} A_\phi(t_1) = \alpha f . \end{equation}
The ratio of scale factors can be expressed as a ratio of temperatures
\begin{equation} \frac{a(t_1)}{a(t_0)} = \frac{T_0}{T_1} , \end{equation}
where the subscript 0 denotes values at the present day, yielding
\begin{equation} \rho_0 = \frac{1}{2}m^2 \alpha^2 f^2 \frac{T_0^3}{T_1^3} . \end{equation}
Using the energy at initial oscillation $T_1 = m$ and the definition of the mass from the previous section, this is simply
\begin{equation} \rho_0 = \frac{1}{2}\alpha^2 \frac{f^3 T_0^3}{\Lambda^2} \sim \frac{f^3 T_0^3}{\Lambda^2} . \end{equation}
So the present PNGB energy density depends on the three relevant temperatures: the temperatures at spontaneous symmetry breaking, at explicit symmetry breaking, and the current CMB temperature of 2.7 K.

\section{Timeline of the PNGB}

\begin{figure}
  \centering
  \includegraphics[width=\textwidth]{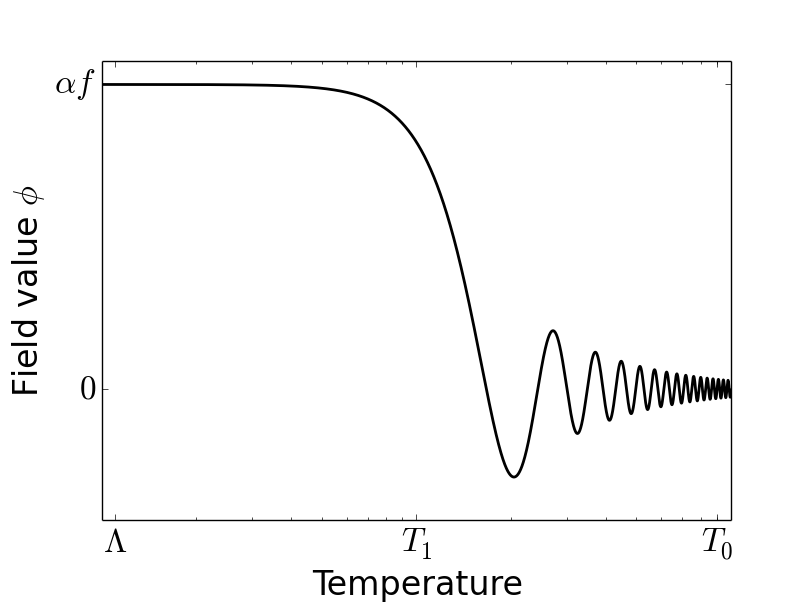}
  \caption{PNGB field value vs. temperature. Temperature is plotted on a log scale.\label{fig:field} }
\end{figure}

\begin{figure}
  \centering
  \includegraphics[width=\textwidth]{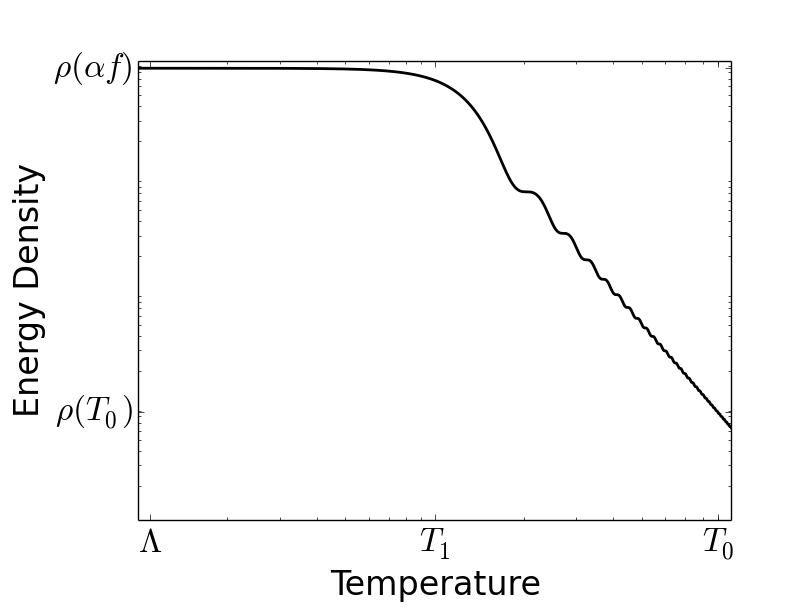}
  \caption{Energy density of the PNGB vs. temperature, plotted on a log-log scale.\label{fig:energy} }
\end{figure}

To summarize the results from the previous sections, we can examine the evolution of the field value in Figure~\ref{fig:field} and energy density in Figure~\ref{fig:energy}. 
\begin{itemize}
\item First the PNGB starts as a Nambu-Goldstone boson when its parent complex scalar field undergoes spontaneous symmetry breaking at the scale $f$. It acquires a vaccum expectation value of $\alpha f$.
\item It retains its vacuum expectation value until it reaches the explicit-symmetry-breaking scale $\Lambda$ and acquires a mass $m=\frac{\Lambda^2}{f}$.
\item Now the PNGB is massive, but its mass is much too small to induce any significant changes. The field ``slowly rolls'' toward its minimum potential energy during this time period since it has a mass, but the change is very small and the majority of its energy density is potential energy, so it acts as vacuum energy with equation of state $w \approx -1$.

\item Finally it reaches the third temperature scale $T_1$, which is set when the mass is the same size as the Hubble parameter $m\sim H$. At this point it begins oscillating. The energy density begins at $\sim m^2f^2 = \Lambda^4$ and evolves with time proportional to $a^{-3}$ until reaching the present temperature $T_0$. Its equation of state is $w \approx 0$, therefore contributing to the matter energy density.
\end{itemize}

\section{The Axion}

The discussion until this point has only referred to a generic PNGB. The above results should roughly hold for a wide range of PNGBs regardless of the physics that generates them, as well as the two specific examples: the axion and quintessence. While they are both PNGBs, they are at different points in their evolution and we will arrive at a few interesting values like the axion-mass lower bound and the quintessence mass.

The axion was originally postulated to solve the strong CP problem.\cite{peccei1996axion} It makes its transition from Nambu-Goldstone boson to PNGB at the QCD scale so the explicit-symmetry-breaking scale for the axion can be identified as
\begin{equation} \Lambda = \Lambda_{QCD} \approx 200 \mbox{ GeV} . \end{equation}
Now the axion's density parameter can be calculated
\begin{equation} \Omega_{axion} = \frac{\rho_0}{\rho_c} , \end{equation}
where $\rho_c$ is the critical density. Finally the requirement that it be less than the dark matter density parameter $\Omega_{axion} < 0.2$, results in an upper limit on $f$ on the order of $10^{12}$ GeV.\cite{sikivie2008axion}

Assuming the axion is the dark matter particle, an approximate bound can be placed on the axion mass using the formula
\begin{equation} m > \frac{\Lambda^2}{f_{max}} = \frac{(200\mbox{ GeV})^2}{10^{21} \mbox{ eV}} = 4 \mu\mbox{eV} , \end{equation}
which sets a lower limit on the axion mass.

\section{Quintessence}

The quintessence field is one possible candidate for dark energy. The dark-energy density is the same order of magnitude as the critical density of the universe at present day, and if the dark energy is of the form of a cosmological constant then its value is unnaturally small at early times. To resolve this ``naturalness problem,'' consider the existence of a PNGB whose explicit-symmetry-breaking scale is very close to the temperature of the universe at present.\cite{ratra1988scalar}$^,$\cite{hill2002pngb}

In the timeline of a PNGB, quintessence is just now entering the slow-roll phase. It has acquired a mass at a scale of the vacuum energy $\Lambda \sim 10^{-3}$ eV, and almost all of its energy is in the form of potential energy, which results in dark energy having a significant contribution to the density today. Furthermore its symmetry-breaking scale is expected to be at the scale of some new unknown physics. A natural scale to postulate for spontaneous symmetry breaking is the Planck scale: $f\sim M_P = 10^{28}$ eV. With these two scales, a naive estimate of the expected mass of the quintessence particle can be obtained\cite{gluscevic2013cmb}
\begin{equation} m = \frac{\Lambda^2}{f} = 10^{-34} \mbox{ eV} . \end{equation}

\section{Birefringence}
\label{sec:birefringence}

Finally, we consider how this relates to birefringence. The Chern-Simons term is of the form
\begin{equation} \mathcal{L}_{CS} = -\frac{\beta \phi}{2M} F^{\mu\nu} \tilde F_{\mu\nu} . \end{equation}
Where $\beta$ is a dimensionless coupling, $\phi$ is the PNGB field, and $M$ is a coupling of mass dimension 1. To see how this could cause a rotation of a linearly polarized photon, consider the modified electromagnetic Lagrangian
\begin{equation} \mathcal{L} = -\frac{1}{4}F^{\mu\nu}F_{\mu\nu} - \frac{\beta \phi}{2M} F^{\mu\nu} \tilde F_{\mu\nu} . \end{equation}
The Euler-Lagrange equations for the field $A_\mu$ are
\begin{equation} \partial^\nu F_{\mu\nu} + \partial_{\nu} \left(\frac{\beta \phi}{M} \varepsilon^{\nu\mu\rho\sigma}F_{\rho\sigma}\right) = 0 . \end{equation}
Since the quantity $\varepsilon^{\nu\mu\rho\sigma}\partial_{\nu}F_{\rho\sigma}$ is identically zero, this equation becomes
\begin{equation} \partial^\nu F_{\mu\nu} + \frac{\beta}{M} \varepsilon^{\mu\rho\sigma\nu}F_{\rho\sigma}\partial_\nu\phi = 0 . \end{equation}
Identifying the components of the field strength tensor with the electric and magnetic fields as
\begin{equation} E^i = -F^{0i} , \end{equation}
\begin{equation} B^i = -\varepsilon^{ijk}F_{jk} , \end{equation}
the two equations take the forms
\begin{equation} \vec{\nabla}\cdot\vec{E} - \frac{\beta}{M}\vec{B}\cdot\vec{\nabla}\phi = 0 , \end{equation}
\begin{equation} \label{eq:rotation} \vec{\nabla}\times\vec{B} - \frac{\partial \vec{E}}{\partial t} + \frac{\beta}{M}\left(\vec{E}\times\vec{\nabla}\phi - \vec{B}\frac{\partial \phi}{\partial t}\right) = 0 . \end{equation} 
For a spatially uniform scalar field, $\phi$ only appears in Eq.~(\ref{eq:rotation}), which is now of the form:
\begin{equation} \vec{\nabla}\times\vec{B} - \frac{\partial \vec{E}}{\partial t} = \frac{\beta}{M}\vec{B}\frac{\partial \phi}{\partial t} . \end{equation}
Therefore if the scalar PNGB field changes over time as a linearly polarized photon travels through the universe, the electric-field vector will change in the direction of the magnetic field. This corresponds to a rotation of the electric field direction. The total rotation angle $\alpha$ can be written in terms of the total change in the PNGB field value\cite{gluscevic2013cmb}
\begin{equation} \alpha = \frac{\beta}{M}\Delta \phi . \end{equation}
Following in the theme of this discussion the magnitude of $M$ should be considered. Since this term deals with $\phi$, it should be at least as big as the spontaneous-symmetry-breaking scale $f$. But recall that the PNGB descended from a theory of a complex scalar field $\psi$, so this coupling likely has a scale associated with some higher energy physics. Again, a plausible scale for this would be the Planck scale $M\sim M_P$.

In order to detect any appreciable rotation angle, the change in the field value of the PNGB must not be more than a few orders of magnitude smaller than $M_P$. The axion's spontaneous-symmetry-breaking scale has an upper bound at around $10^{21}$ eV, so any rotation of photon polarization due to the axion field will could be at largest on the order of $10^{-7}$ rad.

On the other hand, quintessence undergoes spontaneous symmetry breaking closer to the Planck scale. If quintessence exists, we would expect to naturally observe rotation angles on the order of the ratio $\frac{f}{M}$, which could conceivably be as large as order unity.

Put in this context, the axion and quintessence fields should both cause cosmological birefringence, but the magnitudes of these effects may be very different, roughly corresponding to the ratio of their spontaneous-symmetry-breaking scales.

\section{The Standard-Model Extension}
\label{sec:sme}

Until now, the focus has been on one possible mechanism for generating cosmic birefringence, but the SME is a framework to characterize all realistic violations of Lorentz symmetry based on effective field theory, while maintaining other desireable features such as gauge invariance, renormalizability, etc.\cite{colladay1997sme}$^,$\cite{colladay1998sme}$^,$\cite{kostelecky2004sme} .

Within this framework, we can classify potential Lorentz violations in the photon sector with two sets of differential operators $\bm{\hat k}_{AF}$ and $\bm{\hat k}_F$, which characterize CPT-odd and CPT-even violations, respectively.\cite{mewes2002smephotons}$^,$\cite{mewes2009smephotons} They appear in the extended electromagnetic Lagrangian\cite{mewes2007smecmb}
\begin{equation} \mathcal{L}_{SME} = -\frac{1}{4}F_{\mu\nu}F^{\mu\nu} + \frac{1}{2}\epsilon^{\kappa\lambda\mu\nu}A_\lambda\left(\bm{\hat k}_{AF}\right)_\kappa F_{\mu\nu} - \frac{1}{4}F_{\kappa\lambda}\left(\bm{\hat k}_F\right)^{\kappa\lambda\mu\nu}F_{\mu\nu}, \end{equation}
where $A_\mu$ is the vector potential, $F_{\mu\nu}$ is the field strength tensor, and the SME operators are defined as
\begin{equation} \left(\bm{\hat k}_{AF}\right)_\kappa = \sum_{d=\mathrm{odd}} \left(k_{AF}^{(d)}\right)_\kappa^{\alpha_1...\alpha_{(d-3)}}\partial_{\alpha_1}...\partial_{\alpha_{(d-3)}}, \end{equation}
\begin{equation} \left(\bm{\hat k}_F\right)^{\kappa\lambda\mu\nu} = \sum_{d=\mathrm{even}} \left(k_F^{(d)}\right)^{\kappa\lambda\mu\nu\alpha_1...\alpha_{(d-4)}}\partial_{\alpha_1}...\partial_{\alpha_{(d-4)}}. \end{equation}

The effect of including the higher-dimension $d$ terms introduces an energy dependence. Expressing the operators in a spherical-harmonic basis, we can write these possible Lorentz violations as first-order modifications to the photon dispersion relation
\begin{equation} \omega = \left[1-\varsigma^0 \pm \sqrt{\left(\varsigma^1\right)^2 + \left(\varsigma^2\right)^2 + \left(\varsigma^3\right)^2}\right]k, \end{equation}
where the ``$\pm$'' corresponds to the two polarizations and the four $\varsigma^i$ are given as sums over SME parameters\cite{mewes2008smecmb}
\begin{eqnarray}
\varsigma^0 &=& \sum_{djm} \omega^{d-4} \ {}_0Y_{jm}(\bm{\hat n})k^{(d)}_{(I)jm}, \label{eq:sigma0} \\
\varsigma^1\pm i\varsigma^2 &=& \sum_{djm} \omega^{d-4} \ {}_{\pm2}Y_{jm}(\bm{\hat n})\left(k^{(d)}_{(E)jm}\mp ik^{(d)}_{(B)jm}\right), \label{eq:sigma12} \\
\varsigma^3 &=& \sum_{djm} \omega^{d-4} \ {}_0Y_{jm}(\bm{\hat n})k^{(d)}_{(V)jm}, \label{eq:sigma3}
\end{eqnarray}
where $j\leq d-2$ and $\bm{\hat n} = -\bm{\hat p}$ is the line-of-sight direction toward the photon's point of origin. The terms in Eqs.~(\ref{eq:sigma0})~and~(\ref{eq:sigma12}) exist only for even values $d\geq 4$ while those in Eq.~(\ref{eq:sigma3}) exist for odd values $d\geq 3$. It is in terms of the four sets of parameters $k^{(d)}_{(I)jm}, k^{(d)}_{(E)jm}, k^{(d)}_{(B)jm}, k^{(d)}_{(V)jm}$ that we can classify Lorentz violations.\cite{mewes2008smecmb}

\section{Using the CMB for Birefringence Tests}

Now in order to determine which of these infinitely many parameters we can constrain using CMB measurements, we should understand to what each of these four $\varsigma^i$ terms correspond. From the dispersion relation we can see that $\varsigma^0$ is the only one that changes the photon speed by the same amount for both polarizations. This term will not generate birefringence, but, because it contains energy dependence from parameters of dimension $d>4$, its effects can be detected, for example, by measuring arrival times of photons with different frequencies from the same source. However, CMB experiments are not ideal for searching for such an effect, since they may operate only over a single frequency band or a few relatively closely spaced bands. Still, we leave such estimates to a future paper.

The terms $\varsigma^1$ and $\varsigma^2$ characterize CPT-even birefringent effects which mix linear polarization and circular polarization. However the polarization of the CMB, generated primarily by Thomson scattering, is not expected to contain circular polarization and as such CMB experiments typically are not designed to search for circular polarization.

Finally, $\varsigma^3$ characterizes CPT-odd birefringent effects, which result in a rotation of linearly polarized photons without conversion to circular polarization. Measurements of CMB polarization are particularly sensitive to this effect. Unlike other polarized astrophysical sources, like gamma-ray bursts (GRBs) or quasars, the CMB is a well-understood source governed by simple linear physics that allows us to predict the initial polarization state of emitted photons to high accuracy. Combined with the fact that the CMB surface of last scattering has redshift $z\sim1100$, the extraordinarily long propagation distance of CMB photons allows any birefringent effects to accumulate. This is why the CMB is the most sensitive probe of Lorentz violations of this type.\cite{mewes2008smecmb}

Each of these potential Lorentz-violating terms carries energy dependence that increases with the dimension $d$ of the SME parameters. While the CMB is a relatively low-energy source, higher-energy sources like GRBs, pulsars, and blazars will give us much tighter constraints on these higher-dimension parameters. However, for the lowest-dimension terms, CMB measurements are not hampered by this energy dependence, and it is for this reason that we restrict our analysis only to the dimension ($d=3$) coefficients of the SME.

In this case, the change in polarization angle of a linearly polarized photon is\cite{mewes2008smecmb}
\begin{equation} \delta\psi_z = \int_0^z \frac{dz}{(1+z)H_z} \sum_{jm} Y_{jm}(\bm{\hat n})k^{(3)}_{(V)jm}, \end{equation}
where the sum is over $j=0,1$. For a CMB photon, this is approximately
\begin{equation} \label{eq:DeltaPsi} \delta\psi_{\mathrm{CMB}} \approx \left(\frac{3.8^\circ}{10^{-43}\mbox{ GeV}}\right) \sum_{jm}Y_{jm}(\bm{\hat n})k^{(3)}_{(V)jm}. \end{equation} 

\section{POLARBEAR Observations}
\label{sec:polarbear}

We can use observations from the \pb\ experiment\cite{PB1Bmodes} to constrain these dimension $d=3$ SME parameters using its three observational patches. These patches are approximately $3^\circ\times3^\circ$, which is relatively small in the context of the dimension 3 parameters we wish to constrain. Eq.~(\ref{eq:DeltaPsi}) contains simple spherical harmonics up to $j=1$, meaning we are dealing with a monopole term and dipole terms. If we take measurements of a constant rotation angle across one of these sky patches as a measurement of $\delta\psi$ at that particular right ascension and declination, then we can constrain direction-dependent combinations of the four $d=3$ coefficients.

\begin{figure}
  \centering
  \includegraphics[width=\textwidth]{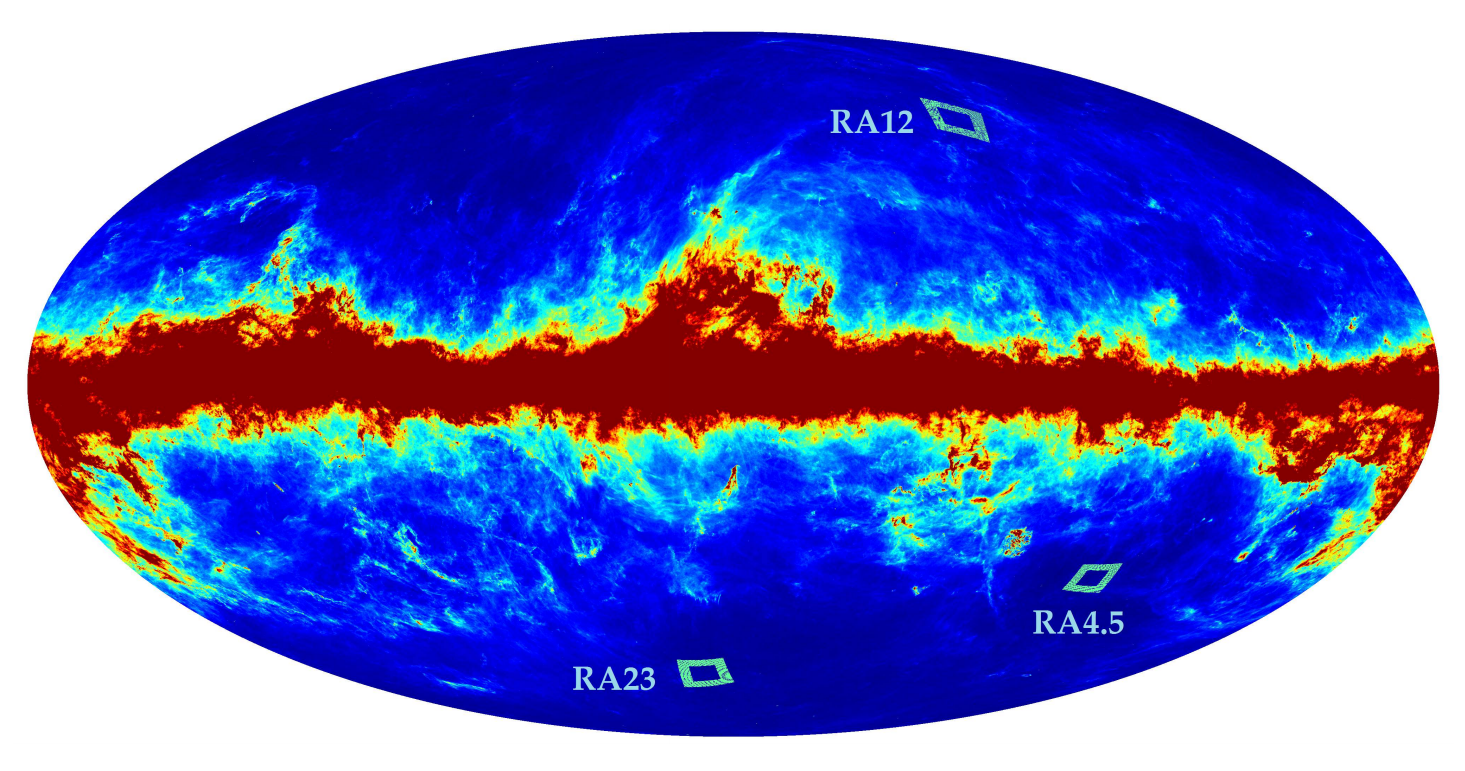}
  \begin{tabular}{lccc}
    \hline\hline
    Patch & RA             & Dec            & Effective Area \\ \hline
    RA4.5 & $04^\mathrm{h}40^\mathrm{m}12^\mathrm{s}$ & $-45^\circ$    & 7.0 deg$^2$ \\
    RA12  & $11^\mathrm{h}53^\mathrm{m}00^\mathrm{s}$ & $ -0^\circ20'$ & 8.7 deg$^2$ \\
    RA23  & $23^\mathrm{h}01^\mathrm{m}48^\mathrm{s}$ & $-32^\circ48'$ & 8.8 deg$^2$
  \end{tabular}
  \caption{The three \pb\ Patches overlaid on a Planck Collaboration full-sky 857 GHz intensity map. \label{fig:PBpatches}\protect\cite{PB1Bmodes}}
\end{figure}

We can see the three patches in Figure~\ref{fig:PBpatches}, along with the values of their RA and Dec. The constraint equations would then be
\begin{eqnarray}
\delta\psi_{\mathrm{ra4.5}} &=& 3.8^\circ\sum_{jm}Y_{jm}(-45^\circ,70^\circ)  \left(\frac{k^{(3)}_{(V)jm}}{10^{-43}\mbox{ GeV}}\right), \\
\delta\psi_{\mathrm{ra12}}  &=& 3.8^\circ\sum_{jm}Y_{jm}(-0.5^\circ,178^\circ)\left(\frac{k^{(3)}_{(V)jm}}{10^{-43}\mbox{ GeV}}\right), \\
\delta\psi_{\mathrm{ra23}}  &=& 3.8^\circ\sum_{jm}Y_{jm}(-33^\circ,345^\circ) \left(\frac{k^{(3)}_{(V)jm}}{10^{-43}\mbox{ GeV}}\right).
\end{eqnarray}

There is one additional complication. During the first season of observations, the \pb\ experiment's absolute angle calibration was obtained by minimizing the $EB$ power spectrum under the assumption of zero overall birefringence.\cite{keating2013calibration} Such an instrumental offset $\alpha'$ will mix the parity-even $E$-mode polarization patterns with the parity-odd $B$-mode polarization patterns to generate spurious $TB$ and $EB$ correlations that are proportional to $\alpha'$. $E$- and $B$-mode maps are rotated by an overall rotation angle to minimize the $EB$ power spectrum in order to remove instrumental miscalibration. This instrumental offset is unfortunately degenerate with a global birefringence angle $\alpha$ meaning that the \pb\ patches cannot constrain the monopole term $k^{(3)}_{(V)00}$ but can still constrain the other coefficients even after this self-calibration procedure by performing the same $EB$ minimization procedure on each of the three patches individually and using the monopole-subtracted rotation angles to constrain the $j=1$ SME coefficients. We leave the calculation of these $j=1$ SME coefficients using data from the \pb\ experiment to a future work.

\section{Conclusion and Outlook}

We have seen now both a theoretical motivation to search for cosmic birefringence and a framework set up by the Standard-Model Extension to use experimental results to place limits on Lorentz-violating effects. Using measurements of the CMB's polarization rotation we can place extremely sensitive constraints on a set of low-dimension SME parameters on the order of $10^{-43}$ GeV.

However, there is still room for improvement. Experiments like \pb\ are in direct need of more accurate calibration measurements. An absolute polarization-angle calibration source would allow \pb\ and other similar CMB experiments to forego self-calibration methods and allow measurements of a global rotation angle offset to constraint isotropic cosmic birefringence as well.\cite{kaufman2016calibration}

The CMB's potential as a probe of parity and Lorentz violation is promising. Through more accurate polarization calibration, or multifrequency analysis, or even a probe of spurious circular polarization we may yet extract even more information from the oldest light in the universe in our search for Lorentz violations in the laws of physics.

\section*{Acknowledgements}

The authors would like to thank Grant Teply for useful discussions and feedback on this paper, and Kevin Crowley for help with the preparation of this paper.

\bibliographystyle{ws-mpla}
\bibliography{references}

\begin{thebibliography}{10}

\bibitem{sikivie2008axion}
P.~{Sikivie}, {\it {Axion Cosmology}}, in {\em Axions\/},  eds. M.~{Kuster},
  G.~{Raffelt} and B.~{Beltr{\'a}n}, Lecture Notes in Physics, Berlin Springer
  Verlag, Vol.~741 (2008).
\newblock p.~19.
\newblock \href{http://arxiv.org/abs/astro-ph/0610440}{{\ttfamily
  astro-ph/0610440}}.

\bibitem{ratra1988scalar}
B.~{Ratra} and P.~J.~E. {Peebles}, {\em \prd} {\bf 37}, 3406 (June 1988).

\bibitem{rosenberg2014axion}
L.~J. {Rosenberg} and G.~{Rybka}, {\em The Review of Particle Physics}  (April
  2014).

\bibitem{ng2001pngb}
S.~C. {Cindy Ng} and D.~L. {Wiltshire}, {\em \prd} {\bf 63},   023503 (January
  2001), \href{http://arxiv.org/abs/astro-ph/0004138}{{\ttfamily
  astro-ph/0004138}}.

\bibitem{ferreira2014axion}
R.~Z. {Ferreira} and M.~S. {Sloth}, {\em Journal of High Energy Physics} {\bf
  12},   139 (December 2014), \href{http://arxiv.org/abs/1409.5799}{{\ttfamily
  arXiv:1409.5799 [hep-ph]}}.

\bibitem{PB1Bmodes}
{The Polarbear Collaboration: P.~A.~R.~Ade}, Y.~{Akiba}, A.~E. {Anthony},
  K.~{Arnold}, M.~{Atlas}, D.~{Barron}, D.~{Boettger}, J.~{Borrill},
  S.~{Chapman}, Y.~{Chinone}, M.~{Dobbs}, T.~{Elleflot}, J.~{Errard},
  G.~{Fabbian}, C.~{Feng}, D.~{Flanigan}, A.~{Gilbert}, W.~{Grainger}, N.~W.
  {Halverson}, M.~{Hasegawa}, K.~{Hattori}, M.~{Hazumi}, W.~L. {Holzapfel},
  Y.~{Hori}, J.~{Howard}, P.~{Hyland}, Y.~{Inoue}, G.~C. {Jaehnig}, A.~H.
  {Jaffe}, B.~{Keating}, Z.~{Kermish}, R.~{Keskitalo}, T.~{Kisner}, M.~{Le
  Jeune}, A.~T. {Lee}, E.~M. {Leitch}, E.~{Linder}, M.~{Lungu}, F.~{Matsuda},
  T.~{Matsumura}, X.~{Meng}, N.~J. {Miller}, H.~{Morii}, S.~{Moyerman}, M.~J.
  {Myers}, M.~{Navaroli}, H.~{Nishino}, A.~{Orlando}, H.~{Paar}, J.~{Peloton},
  D.~{Poletti}, E.~{Quealy}, G.~{Rebeiz}, C.~L. {Reichardt}, P.~L. {Richards},
  C.~{Ross}, I.~{Schanning}, D.~E. {Schenck}, B.~D. {Sherwin}, A.~{Shimizu},
  C.~{Shimmin}, M.~{Shimon}, P.~{Siritanasak}, G.~{Smecher}, H.~{Spieler},
  N.~{Stebor}, B.~{Steinbach}, R.~{Stompor}, A.~{Suzuki}, S.~{Takakura},
  T.~{Tomaru}, B.~{Wilson}, A.~{Yadav} and O.~{Zahn}, {\em \apj} {\bf 794},
  171 (October 2014), \href{http://arxiv.org/abs/1403.2369}{{\ttfamily
  arXiv:1403.2369}}.

\bibitem{peccei1996axion}
R.~D. {Peccei}, {\em Journal of Korean Physical Society} {\bf 29},   199
  (September 1996), \href{http://arxiv.org/abs/hep-ph/9606475}{{\ttfamily
  hep-ph/9606475}}.

\bibitem{hill2002pngb}
C.~T. {Hill} and A.~K. {Leibovich}, {\em \prd} {\bf 66},   075010 (October
  2002), \href{http://arxiv.org/abs/hep-ph/0205237}{{\ttfamily
  hep-ph/0205237}}.

\bibitem{gluscevic2013cmb}
V.~{Gluscevic}, {CMB as a Probe of New Physics and Old Times}, PhD thesis,
  California Institute of Technology, (2013).

\bibitem{colladay1997sme}
D.~{Colladay} and V.~A. {Kosteleck{\'y}}, {\em \prd} {\bf 55}, 6760 (June
  1997), \href{http://arxiv.org/abs/hep-ph/9703464}{{\ttfamily
  hep-ph/9703464}}.

\bibitem{colladay1998sme}
D.~{Colladay} and V.~A. {Kosteleck{\'y}}, {\em \prd} {\bf 58},   116002
  (December 1998), \href{http://arxiv.org/abs/hep-ph/9809521}{{\ttfamily
  hep-ph/9809521}}.

\bibitem{kostelecky2004sme}
V.~A. {Kosteleck{\'y}}, {\em \prd} {\bf 69},   105009 (May 2004),
  \href{http://arxiv.org/abs/hep-th/0312310}{{\ttfamily hep-th/0312310}}.

\bibitem{mewes2002smephotons}
V.~A. {Kosteleck{\'y}} and M.~{Mewes}, {\em \prd} {\bf 66},   056005 (September
  2002), \href{http://arxiv.org/abs/hep-ph/0205211}{{\ttfamily
  hep-ph/0205211}}.

\bibitem{mewes2009smephotons}
V.~A. {Kosteleck{\'y}} and M.~{Mewes}, {\em \prd} {\bf 80},   015020 (July
  2009), \href{http://arxiv.org/abs/0905.0031}{{\ttfamily arXiv:0905.0031
  [hep-ph]}}.

\bibitem{mewes2007smecmb}
V.~A. {Kosteleck{\'y}} and M.~{Mewes}, {\em Physical Review Letters} {\bf 99},
   011601 (July 2007), \href{http://arxiv.org/abs/astro-ph/0702379}{{\ttfamily
  astro-ph/0702379}}.

\bibitem{mewes2008smecmb}
V.~A. {Kosteleck{\'y}} and M.~{Mewes}, {\em \apjl} {\bf 689}, L1 (December
  2008), \href{http://arxiv.org/abs/0809.2846}{{\ttfamily arXiv:0809.2846}}.

\bibitem{keating2013calibration}
B.~G. {Keating}, M.~{Shimon} and A.~P.~S. {Yadav}, {\em \apjl} {\bf 762},   L23
  (January 2013), \href{http://arxiv.org/abs/1211.5734}{{\ttfamily
  arXiv:1211.5734 [astro-ph.CO]}}.

\bibitem{kaufman2016calibration}
J.~P. {Kaufman}, B.~G. {Keating} and B.~R. {Johnson}, {\em \mnras} {\bf 455},
  1981 (January 2016).

\end{thebibliography}

\end{document}